\documentclass[12pt]{article}
\usepackage{bbm}
\usepackage{graphicx}
\usepackage{amsmath}
\usepackage{amssymb}
\usepackage{caption2}
\begin{document}
\bibliographystyle{prsty}
\begin{center}
{\large {\bf \sc{  Final-state interactions in the decays $B^0\to
\chi_{c0}K^{*0}$ and $B^+
\to \chi_{c0}K^{*+}$  }}} \\[2mm]
Zhi-Gang Wang \footnote{ E-mail,wangzgyiti@yahoo.com.cn.  }    \\
 Department of Physics, North China Electric Power University, Baoding 071003, P. R. China
\end{center}

\begin{abstract}
In this article, we study the  final-state rescattering effects in
the decays $B^0\to \chi_{c0}K^{*0}$ and $B^+ \to \chi_{c0}K^{*+}$,
and observe the corrections are zero in the $SU(3)$ limit,  which is
warranted by the heavy quark symmetry. It is difficult to
   accommodate the experimental data without fine-tuning.
\end{abstract}


\section{Introduction}

Recently, the Babar collaboration reported the observation of the
decay $B^0\to \chi_{c0}K^{*0}$ as well as evidence of the decay $B^+
\to \chi_{c0}K^{*+}$ with an $8.9$ and a $3.6$ standard deviation
significance respectively \cite{BABAR0808}. The measured branching
fractions are $ Br(B^0  \rightarrow \chi_{c0}K^{*0}) = (1.7 \pm 0.3
\pm 0.2) \times 10^{-4}$ and $Br(B^+ \rightarrow \chi_{c0}K^{*+}) =
(1.4 \pm 0.5 \pm 0.2) \times 10 ^{-4}$. They also obtained  the
upper limit  $ Br(B^+ \rightarrow \chi_{c0}K^{*+}) < 2.1 \times
10^{-4}$ at the $90\%$ confidence level. The decays take place
through the process $b \to sc \bar{c}$ (or more precise $\bar{b} \to
\bar{s} c\bar{c}$, they relate with each other by  charge
conjunction, in this article, we calculate the amplitudes for the
process $b \to sc \bar{c}$, then take  charge conjunction to obtain
the branching fractions.) at the quark-level. The quantitative
understanding of those decays depends on our knowledge about the
nonperturbative hadronic matrix elements of the operators entering
the effective weak Hamiltonian. The factorizable contributions in
the decays $B^0\rightarrow \chi_{c0}K^{*0}$ and $B^+ \rightarrow
\chi_{c0}K^{*+}$ are zero, we have to resort to special  mechanism
to overcome the difficulty.

Final-state interactions play an important role in the hadronic
$B$-decays,  the  color-suppressed neutral modes such as $B^0 \to
D^0 \pi^0, \pi^0\pi^0, \rho^0\pi^0, K^0\pi^0$ are  enhanced
substantially by the long-distance rescattering effects \cite{CHHY}.
In Refs.\cite{ColangeloFSI1,ColangeloFSI2}, the authors study  the
rescattering effects of the  intermediate charmed mesons for the
decays $B^- \to \chi_{c0}K^- , h_cK^- $,
  and observe the final-state interactions can
lead to larger branching fractions to account the experimental data.
The factorizable amplitude in the decay  $B^0\to \eta_cK^*$ is too
small to accommodate the experimental data \cite{BABAR0804}, and the
effects of the final-state interactions can smear the discrepancy
\cite{Wang0804}.
 It is intersecting to study the effects of the final-state
interactions in the decays $B^0\rightarrow \chi_{c0}K^{*0}$ and $B^+
\rightarrow \chi_{c0}K^{*+}$.

The article is arranged as: in Section 2, we study  the final-state
rescattering effects in the decays $B^0\rightarrow \chi_{c0}K^{*0}$
and $B^+ \rightarrow \chi_{c0}K^{*+}$; in Section 3, the numerical
result and discussion; and Section 4 is reserved for conclusion.

\section{Final-state rescattering effects in the decays  $B^0\rightarrow \chi_{c0}K^{*0}$ and $B^+
\rightarrow \chi_{c0}K^{*+}$}

The effective weak Hamiltonian for the  decay modes $b\rightarrow s
c \bar{c}$ can be written as (for detailed discussion of the
effective weak Hamiltonian, one can consult Ref.\cite{Buras})
\begin{equation}
H_w = \frac{G_F}{\sqrt{2}}\left\{V_{c b} V_{c s}^* \left[ C_1(\mu)
{\cal O}_1(\mu) +  C_2(\mu) {\cal O}_2(\mu) \right]+h.c. \right\}
\, ,
\end{equation}
where the $V_{ij}$'s are the CKM matrix elements, the $C_i$'s are
the Wilson coefficients calculated at the renormalization scale $\mu
\sim O(m_b)$ and the relevant operators ${\cal O}_i$  are given by
 \begin{eqnarray}
 {\cal O}_1&=&(\overline{s}_{\alpha} b_{\alpha})_{V-A}
(\overline{c}_{\beta} c_{\beta})_{V-A}\, , \nonumber\\
 {\cal O}_2&=&(\overline{s}_{\alpha} b_{\beta})_{V-A}
(\overline{c}_{\beta} c_{\alpha})_{V-A}\,,
  \end{eqnarray}
 here $\alpha$ and $\beta$ are color indexes. From the effective weak Hamiltonian $H_w$,
 we can see that
  the factorizable
amplitudes are zero.

The decays $B^0 \to DD_s$, $DD_s^*$, $D^*D_s$, $D^*D_s^*$  are color
enhanced due to the large Wilson coefficient $C_2$ \footnote{The
corresponding decays for the $B^+$ can be studied in the same way,
in the following, we only present the technical details for the
$B^0$ decays. },
\begin{eqnarray}
\langle D_s(q)D^*(p)| H_w|B(P) \rangle &=&2\widetilde{a}_2 P\cdot\epsilon^*(p) f_{D_s}M_{D^*}A_0(q^2)\, ,\nonumber \\
\langle D_s^*(q)D(p)| H_w|B(P) \rangle &=&2\widetilde{a}_2
P\cdot\epsilon^*(q) f_{D_s^*}M_{D^*_s}F_1(q^2) \, ,\nonumber \\
\langle D_s(q)D(p)| H_w|B(P) \rangle &=&\widetilde{a}_2 (M_D^2-M_B^2) f_{D_s}F_0(q^2) \, ,\nonumber \\
\langle D_s^*(q)D^*(p)| H_w|B(P) \rangle &=&\widetilde{a}_2
f_{D^*_s}M_{D^*_s}\left[
\frac{2\epsilon^{\mu\nu\alpha\beta}\epsilon_\mu^*(q)
\epsilon_\nu^*(p)P_\alpha p_\beta V(q^2)}{M_B+M_{D^*}} -
\epsilon^*(q) \cdot
\epsilon^*(p)\right.  \nonumber \\
&&\left.(M_B+M_{D^*})A_1(q^2)+\frac{2P\cdot \epsilon^*(q)
\epsilon^*(p) \cdot q A_2(q^2)}{M_B+M_{D^*}} \right] \, ,
\end{eqnarray}
where $\widetilde{a}_2=\frac{G_F}{\sqrt{2}} V_{c b} V_{c s}^*
(C_2+\frac{C_1}{3})$,  the $f_D$, $f_{D^*}$, $f_{D_s}$, $f_{D^*_s}$
are the weak decay constants, and the $A_0(q^2)$, $A_1(q^2)$,
$A_2(q^2)$,  $V(q^2)$, $F_0(q^2)$, $F_1(q^2)$ are the weak
form-factors defined by \cite{Form1,Form2},
\begin{eqnarray}
\langle D(p)| \overline{s} \gamma_{\mu}(1-\gamma_5) b | B(P) \rangle
&=& (P + p)_{\mu}  F_{1}(q^2) - \frac{ M_B^2 - M_D^2 }{q^2} q_{\mu}
[F_{1}(q^2) - F_0(q^2)]\, , \nonumber\\
\langle D^*(p)| \overline{s} \gamma_{\mu}(1-\gamma_5) b | B(P)
\rangle&=&{\epsilon_\mu}^{\nu\alpha\beta}\epsilon^*_\nu P_\alpha
p_\beta
\frac{2V(q^2)}{M_B+M_{V}} -\frac{2M_{V}q\cdot\epsilon^*}{q^2}q_\mu A_0(q^2)\nonumber\\
&&-\left(\epsilon_\mu^*-\frac{q\cdot\epsilon^*}{q^2}q_\mu\right)(M_B+M_{V})A_1(q^2)+\nonumber\\
&&\left[(P+p)_\mu-\frac{M_B^2-M_{V}^2}{q^2}q_\mu\right]q\cdot\epsilon^*
\frac{A_2(q^2)}{M_B+M_{V}} \, ,
\end{eqnarray}
and the $\epsilon_\mu$ is the polarization vector of the vector
meson, $q_\mu=P_\mu-p_\mu$.

 The decays  $B^0\rightarrow \chi_{c0}K^{*0}$ and $B^+
\rightarrow \chi_{c0}K^{*+}$ can take place through the decay
cascades $B \to DD_s$, $DD_s^*$, $D^*D_s$, $D^*D_s^*$ $\to \chi_{c0}
K^*$, the rescattering amplitudes of  $DD_s$, $DD_s^*$, $D^*D_s$,
$D^*D_s^*$ $\to \chi_{c0} K^*$ may play an important role.

The final-state interactions can be described  by the following
effective lagrangians,
\begin{eqnarray}
\mathcal{L}_{\chi_{c0} DD}&=& g_{\chi_c DD} \chi_{c0}  DD^{\dagger}\, ,\\
\mathcal{L}_{\chi_{c0} D^*D^*}&=& g_{\chi_c D^*D^*} \chi_{c0}  D^* \cdot D^{*\dagger}\, ,\\
\mathcal{L}_{DDV}&=&-ig_{DDV}D_{i}^{\dagger}{\stackrel{\leftrightarrow}{\partial}}
_{\mu}D^{j}(\mathbb{V}^{\mu})^{i}_{j} \, ,\\
\mathcal{L}_{D^*DV}&=&-2f_{D^{*}DV}\varepsilon^{\mu\nu\alpha\beta}(\partial_{\mu}\mathbb{V}_{\nu})^{i}_{j}
\left[ D_{i}^{\dagger}{\stackrel{\leftrightarrow}{\partial}}_{\alpha}D_\beta^{* j}-D^{*\dagger}_{\beta i}{\stackrel{\leftrightarrow}{\partial}}_{\alpha}D^{j}\right]\, ,\\
\mathcal{L}_{D^*D^*V}&=&ig_
{D^{*}D^{*}V}D^{*\nu\dagger}_{i}{\stackrel{\leftrightarrow}{\partial}}_{\mu}
D^{*j}_{\nu}(\mathbb{V}^{\mu})^{i}_{j}+4if_{D^{*}D^{*}V}D^{*\dagger}_{i\mu}(\partial^{\mu}\mathbb{V}^{\nu}
-\partial^{\nu}\mathbb{V}^{\mu})^{i}_{j}D^{*j}_{\nu}\, ,
\end{eqnarray}
where the indexes $i,j$  stand for  the flavors of the light quarks,
$D^{(*)}$=$(\bar{D}^{(*)0}$, $D^{(*)-}$, $D_{s}^{(*)-})^T$, and
$\mathbb{V}$ is the $3\times 3$ matrix for the nonet vector mesons,
\begin{eqnarray}
\mathbb{V}&=&\left(\begin{array}{ccc}
\frac{\rho^{0}}{\sqrt{2}}+\frac{\omega}{\sqrt{2}}&\rho^{+}&K^{*+}\\
\rho^{-}&-\frac{\rho^{0}}{\sqrt{2}}+\frac{\omega}{\sqrt{2}}&
K^{*0}\\
K^{*-} &\bar{K}^{*0}&\phi
\end{array}\right).
\end{eqnarray}
 The lagrangians
$\mathcal{L}_{DDV}$, $\mathcal{L}_{D^*DV}$ and
$\mathcal{L}_{D^*D^*V}$ are taken from Ref.\cite{CHHY}, and the
$\mathcal{L}_{\chi_{c0} D^*D}$ and $\mathcal{L}_{\chi_{c0} D^*D^*}$
are constructed from the   heavy quark theory  in this article.

The rescattering effects can be taken into account  by eight Feynman
diagrams,  see Fig.1. We  calculate the absorptive parts (or
imaginary parts) of the rescattering amplitudes $\textbf{Abs}(i)$ by
the Cutkosky rule,
\begin{eqnarray}
\textbf{Abs}(i)&=&\frac{1}{2}\int \frac{d^3\vec{p}_1}{(2\pi)^32E_1}
\int \frac{d^3\vec{p}_2}{(2\pi)^32E_2} (2\pi)^4
\delta^4(P-p_1-p_2)\mathcal {T}^i_{B\to int} \mathcal {T}^i_{int\to
\chi_{c0}K^*}
 \, , \nonumber \\
\end{eqnarray}
where the amplitudes $\mathcal {T}^i_{B\to int}$ stand for the
corresponding factorizable contributions presented in Eq.(3), and
the rescattering  amplitudes $\mathcal {T}^i_{int\to \chi_{c0}K^*}$
are given by
\begin{eqnarray}
\mathcal {T}^a_{DD_s\to \chi_{c0}K^*}&=& -2ig_{DDV}p_2 \cdot
\epsilon^*(p_4) \frac{1}{t-M_D^2}\mathcal {F}^2(t)ig_{\chi_{c0} DD} \, ,\nonumber\\
\mathcal {T}^b_{D_sD\to \chi_{c0}K^*}&=& 2ig_{DDV}p_2 \cdot
\epsilon^*(p_4) \frac{1}{t-M_{D_s}^2}\mathcal {F}^2(t)ig_{\chi_{c0} DD}\, , \nonumber\\
\mathcal {T}^c_{D^*D_s\to \chi_{c0}K^*}&=& 4if_{D^*DV}
\epsilon^{\mu\nu\alpha\beta}p_{4\mu}
\epsilon^*_\nu(p_4)p_{2\alpha}\epsilon_\beta(q)
 \frac{1}{t-M_{D^*}^2}\mathcal {F}^2(t)ig_{\chi_{c0} D^*D^*} \epsilon^*(q) \cdot \epsilon(p_1)\, ,\nonumber\\
 \mathcal {T}^d_{D^*_sD\to \chi_{c0}K^*}&=& 4if_{D^*DV}
\epsilon^{\mu\nu\alpha\beta}p_{4\mu}
\epsilon^*_\nu(p_4)p_{2\alpha}\epsilon_\beta(q)
 \frac{1}{t-M_{D^*_s}^2}\mathcal {F}^2(t)ig_{\chi_{c0} D^*D^*} \epsilon^*(q) \cdot \epsilon(p_1)\, ,\nonumber\\
 \mathcal {T}^e_{DD^*_s\to \chi_{c0}K^*}&=& -4if_{D^*DV}
\epsilon^{\mu\nu\alpha\beta}p_{4\mu}
\epsilon^*_\nu(p_4)p_{2\alpha}\epsilon_\beta(p_2)
 \frac{1}{t-M_{D}^2}\mathcal {F}^2(t)ig_{\chi_{c0} DD} \, ,\nonumber\\
 \mathcal {T}^f_{D_sD^*\to \chi_{c0}K^*}&=& -4if_{D^*DV}
\epsilon^{\mu\nu\alpha\beta}p_{4\mu}
\epsilon^*_\nu(p_4)p_{2\alpha}\epsilon_\beta(p_2)
 \frac{1}{t-M_{D_s}^2}\mathcal {F}^2(t)ig_{\chi_{c0} DD} \, ,\nonumber\\
\mathcal {T}^g_{D^*D^*_s\to \chi_{c0}K^*}&=& \left\{4if_{D^*D^*V}
\left[ p_4 \cdot \epsilon(p_2) \epsilon^*(p_4)\cdot \epsilon(q)-p_4
\cdot \epsilon(q) \epsilon^*(p_4)\cdot \epsilon(p_2) \right]+
\right. \nonumber \\
&&\left. 2ig_{D^*D^*V}\epsilon^*(p_4) \cdot p_2 \epsilon(q) \cdot
\epsilon(p_2)\right\}
 \frac{1}{t-M_{D^*}^2}\mathcal {F}^2(t)ig_{\chi_{c0} D^*D^*}\epsilon^*(q) \cdot \epsilon(p_1) \, ,\nonumber\\
 \mathcal {T}^h_{D^*_sD^*\to \chi_{c0}K^*}&=& \left\{-4if_{D^*D^*V}
\left[ p_4 \cdot \epsilon(p_2) \epsilon^*(p_4)\cdot \epsilon(q)-p_4
\cdot \epsilon(q) \epsilon^*(p_4)\cdot \epsilon(p_2) \right]-
\right. \nonumber \\
&&\left. 2ig_{D^*D^*V}\epsilon^*(p_4) \cdot p_2 \epsilon(q) \cdot
\epsilon(p_2)\right\}
 \frac{1}{t-M_{D_s^*}^2}\mathcal {F}^2(t)ig_{\chi_{c0} D^*D^*}\epsilon^*(q) \cdot \epsilon(p_1) \, ,\nonumber\\
\end{eqnarray}
where $t=q^2$, $q=p_1-p_3=p_4-p_2$, and the $\epsilon_\mu$ is the
polarization vector of the corresponding  vector meson $V$,
$\epsilon^*_\alpha(q) \epsilon_\beta(q) \to
-g_{\alpha\beta}+\frac{q_\alpha q_\beta}{M_V^2}$. The $p_1$, $p_2$,
$p_3$ and $p_4$ stand for the momenta of the mesons $D$, $D_s$,
$\chi_{c0}$ and $K^*$ respectively in the amplitude $\mathcal
{T}^{a}_{DD_s \to \chi_{c0}K^*}$; the momenta in other amplitudes
can be understood analogously.  The off-shell effects of the
$t$-channel exchanged mesons $D$, $D^*$, $D_s$ and $D_s^*$ are taken
into account  by introducing a monopole form-factor \cite{CHHY},
\begin{eqnarray}
\mathcal{F}(M_{i},t)=\frac{\Lambda_i^{2}-M_{i}^2 }{\Lambda_i^{2}-t},
\end{eqnarray}
and the cutoff $\Lambda_i$ are parameterized as
$\Lambda_{i}=M_{i}+\alpha \Lambda_{\rm{QCD}} $, where $\alpha$ is a
free parameter and $\Lambda_{\rm{QCD}}=0.225\rm{GeV}$. In fact, the
$g_s\mathcal{F}(M_{i},t)$ are the momentum dependent strong coupling
constants, we can vary the parameter $\alpha$ to change the
effective strong couplings,  here we use the notation $g_s$ to
denote all the strong coupling constants.

The dispersive parts (or real parts) of the rescattering amplitudes
can be obtained via the dispersion relation,
\begin{eqnarray}
 \textbf{Dis}(i) (M_B^2) = {1 \over \pi} \textbf{P}
\int_{s_{th}}^{\infty} { \textbf{Abs}(i) (s^\prime) \over s^\prime -
M_B^2} d s^\prime \, ,
\end{eqnarray}
where the thresholds $s_{th}$ are given by $s_{th}=(
M_D+M_{D^*_s})^2$ , $( M_{D^*}+M_{D_s})^2$, $( M_D+M_{D_s})^2$, $(
M_{D^*}+M_{D^*_s})^2$ for any specific diagram.

\section{Numerical result and discussions}
In the flavor $SU(3)$ limit, there are strong cancelation among the
rescattering amplitudes, $\mathcal {T}^a_{DD_s\to
\chi_{c0}K^*}+\mathcal {T}^b_{D_sD\to \chi_{c0}K^*}=0$ and $
\mathcal {T}^g_{D^*D^*_s\to \chi_{c0}K^*}+\mathcal
{T}^h_{D^*_sD^*\to \chi_{c0}K^*}=0$. From the Particle Data Group,
$M_{D}=1.87\,\rm{GeV}$, $M_{D_s}=1.97\,\rm{GeV}$,
$M_{D^*}=2.010\,\rm{GeV}$ and $M_{D_s^*}=2.112\,\rm{GeV}$
\cite{PDG}, we can see that the $SU(3)$ breaking effects are small.
However,  the experimental data from the CLEO collaboration,
$f_D=222.6\pm16.7^{+2.8}_{-3.4}\,\,\rm{MeV}$
\cite{decayCP1,decayCP2} and  $f_{D_s}=(0.274\pm0.013)\,\rm{GeV}$
\cite{decayCPs}
  show that  the $SU(3)$ breaking effects are rather large,
$\frac{f_{D_s}}{f_D}=1.23$, while most of theoretical calculations
indicate $\frac{f_{D_s}}{f_D}\approx 1.1$,  the discrepancy maybe
indicate new physics beyond the standard model \cite{Narison0807}.
If we take into account the small $SU(3)$ breaking effects, the
rescattering amplitudes $\mathcal {T}^i$ ($i=a,b,g,h$) have
contributions.

For the rescattering amplitudes $\mathcal {T}^i$,  $i=c,d,e,f$,
there are no cancelation. We carry out the integrals formally,
\begin{eqnarray}
\textbf{Abs}(i) &=&\int \frac{d^3\vec{p}_1}{2E_1} \int
\frac{d^3\vec{p}_2}{2E_2} \delta^4(P-p_1-p_2) f(p_1,p_2,p_3,p_4)
\epsilon^{\mu\nu\alpha\beta}p_{4\mu}
\epsilon^*_\nu(p_4)p_{2\alpha}p_{1\beta} \nonumber \\
&=&  \epsilon^{\mu\nu\alpha\beta}p_{4\mu}
\epsilon^*_\nu(p_4) \left[Ag_{\alpha\beta}+Bp_{3\alpha}p_{3\beta} +Cp_{3\alpha}p_{4\beta}+Dp_{4\alpha}p_{4\beta}\right]\nonumber \\
&=&0 \, ,
\end{eqnarray}
where we have introduced the formal notations $f(p_1,p_2,p_3,p_4)$
and $A$, $B$, $C$, $D$ (scalar coefficients), they have no
contributions.

The above conclusion and the following discussion also hold for the
decay $B^+ \to \chi_{c0}K^{*+}$.

The strong coupling constants $f_{D^*DV}$, $f_{D^*D^*V}$, $g_{DDV}$,
 $g_{D^*D^*V}$, $g_{\chi_c DD}$ and $g_{\chi_c D^*D^*}$ are not free parameters.
  In the heavy quark limit, the strong coupling
constants $f_{D^*DV}$, $f_{D^*D^*V}$, $g_{DDV}$ and $g_{D^*D^*V}$
can be related to the basic parameters  $\lambda$ and $\beta$ in the
heavy quark effective Lagrangian (one can consult Ref.\cite{HQEFT97}
for the heavy  quark effective lagrangian and relevant parameters,
we neglect them for simplicity),
\begin{eqnarray}
 f_{D^*DV}&=&\frac{f_{D^*D^*V}}{M_{D^*}}=\frac{\lambda g_V}{\sqrt{2}}\, ,\nonumber\\
g_{DDV}&=&g_{D^*D^*V}=\frac{\beta g_V}{\sqrt{2}}\, ,
\end{eqnarray}
where  $g_V=5.8$  from the vector meson dominance theory
\cite{VMDgV}; we can also calculate them with the light-cone QCD sum
rules \cite{WangCCT1,WangCCT2}. The strong coupling constants
$g_{\chi_c DD}$ and $g_{\chi_c D^*D^*}$ can be estimated with the
universal Isgur-Wise form-factor at zero recoil $\xi(1)$ and the
assumption of dominance of the intermediate $\chi_{c0}$ meson for
the scalar heavy quark current $\overline{c} c$ \cite{HeavyQuark},
\begin{eqnarray}
g_{\chi_c DD}&=&\frac{2M_DM_{\chi_{c0}}}{f_{\chi_{c0}}} \, , \nonumber \\
g_{\chi_c D^*D^*}&=&-\frac{2M_{D^*}M_{\chi_{c0}}}{f_{\chi_{c0}}} \,
,
\end{eqnarray}
where the decay constant $f_{\chi_{c0}}$ is defined by
$f_{\chi_{c0}}M_{\chi_{c0}}=\langle 0|
\bar{c}(0)c(0)|\chi_{c0}\rangle$.

The only free parameter is the $\alpha$ in the off-shell
form-factors $\mathcal{F}(M_{i},t)$. We may take into account the
experimental data from the Babar collaboration with fine-tuning of
the momentum dependent strong coupling constants
$g_s\mathcal{F}(M_{i},t)$. We will not resort to the fine-tuning
mechanism.

In fact, the $t$-channel rescattering amplitudes $DD_s$, $DD_s^*$,
$D^*D_s$, $D^*D_s^*$ $\to \chi_{c0} K^*$ can be described by the
collective strong coupling constant $\mathbbm{g}$,
$\mathbbm{g}=g_s^2\mathcal{F}^2(M_{i},t)\frac{1}{t-M^2_i}$.
 At the level of  quark-gluon degrees of freedom, the rescattering occur
 through
$c\bar{q}_1+\bar{c}q_2 \to \chi_{c0} +V$. In the heavy quark limit,
the heavy quarks decouple from the light degrees of freedom, the
$\bar{q}_1 q_2$ rearrange to the vector meson $V$ (or  pseudoscalar
meson $P$), conservation of the heavy quark spin warrants the
$\bar{c}c$ pair rearrange to the $J/\psi$ or $\eta_c$, not the
$\chi_{c0}$, because additional relative $P$-wave between the
$\bar{c}c$  pair is required to form the $\chi_{c0}$, $\chi_{c1}$
and $\chi_{c2}$. It is not unexpected that  the total rescattering
amplitudes are nearly zero, the small discrepancy  due to the
$SU(3)$ breaking effects and the $c$ quark is not heavy enough. This
case  is  contrary to the $^3P_0$ model \cite{3P01,3P02}, where the
$\bar{q}q$ pairs with the quantum numbers $^3P_0$ are created from
the QCD vacuum, the relative $P$-wave between the $\bar{c}c$ pair is
canceled out with the relative $P$-wave between the  $\bar{q}q$
pair, the decays $\chi_{c0}+\bar{q}q \to D\bar{D}, D^*\bar{D}^*$ can
occur, if they are kinetically allowed, see the effective
lagrangians in Eqs.(5-6). The final state interactions may play an
important role in the precesses with the final states
$\chi_{cJ}+S,A$ and $J/\psi,\eta_c+P,V$, where the $S$, $P$, $V$ and
$A$ stand for the scalar, pseudoscalar, vector and axial-vector
mesons respectively.

 \section{Conclusion}
In this article, we study the  final-state rescattering effects in
the decays $B^0\to \chi_{c0}K^{*0}$ and $B^+ \to \chi_{c0}K^{*+}$,
and observe the corrections are zero in the $SU(3)$ limit, which is
warranted by the heavy quark symmetry. It is difficult to
   accommodate the experimental data without fine-tuning. The final state
   interactions may play an important role in the
decays  $B \to \chi_{cJ}+S,A; J/\psi,\eta_c+P,V$.

\section*{Acknowledgments}
This  work is supported by National Natural Science Foundation,
Grant Number  10775051, and Program for New Century Excellent
Talents in University, Grant Number NCET-07-0282.

\newpage
\begin{figure}
\centering
  \includegraphics[totalheight=20cm,width=17cm]{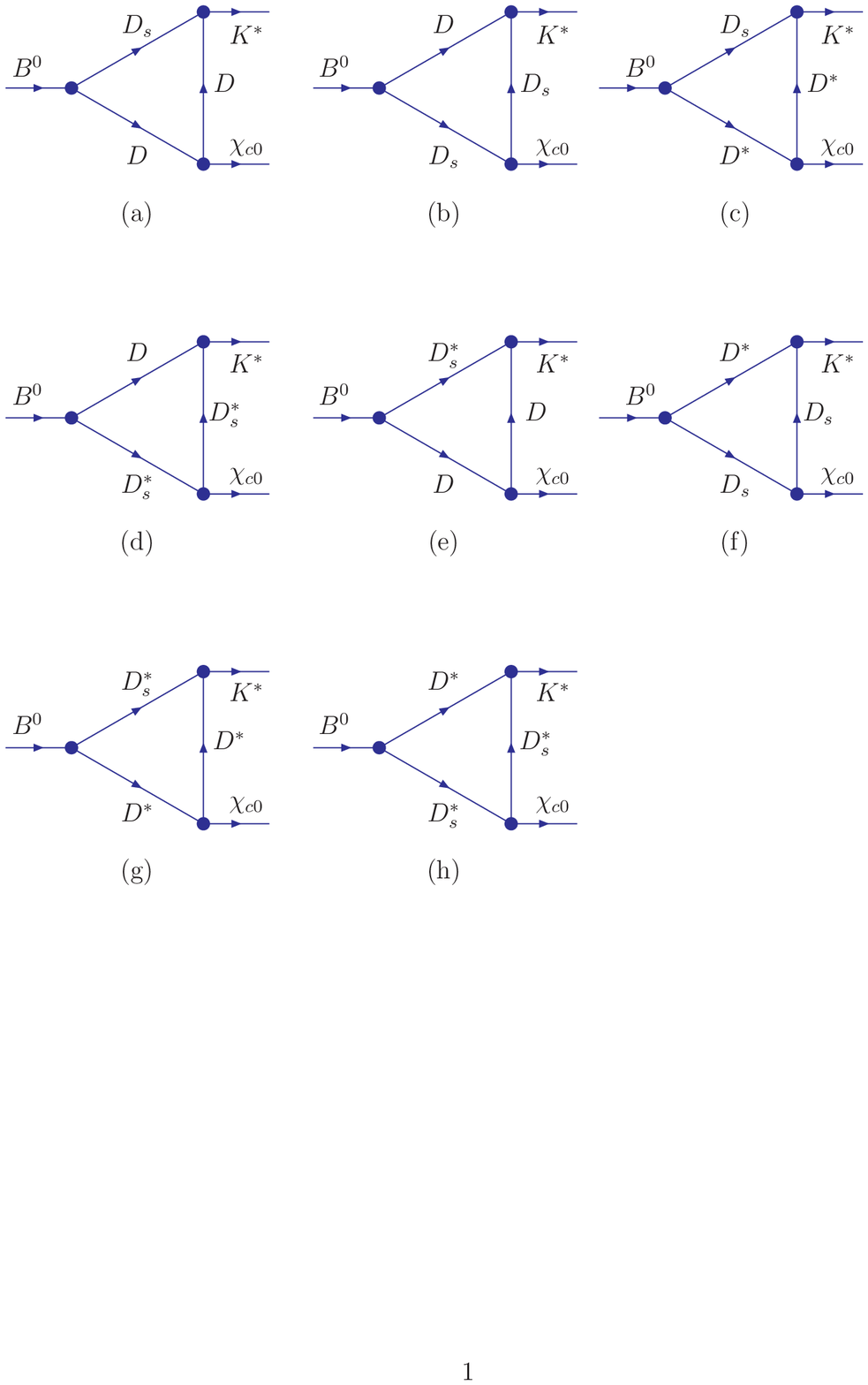}
      \caption{The Feynman diagrams for the final-state interactions. }
\end{figure}


\begin{thebibliography}{99}


 \bibitem{BABAR0808} B. Aubert et al, arXiv:0808.1487.

\bibitem{CHHY}   H. Y. Cheng, C. K. Chua and A. Soni, Phys. Rev. {\bf D71} (2005) 014030.


\bibitem{ColangeloFSI1} P. Colangelo, F. De Fazio and T. N. Pham, Phys. Lett. {\bf B542} (2002) 71.

\bibitem{ColangeloFSI2} P. Colangelo, F. De Fazio and T. N. Pham, Phys. Rev. {\bf D69} (2004) 054023.

\bibitem{BABAR0804} B. Aubert et al, Phys. Rev. {\bf  D78} (2008) 012006.

\bibitem{Wang0804} Z. G. Wang, arXiv:0804.4227.


\bibitem{Buras}  G. Buchalla, A. J. Buras and M. E. Lautenbacher,
Rev. Mod. Phys. {\bf 68} (1996) 1125.


\bibitem{Form1} M. Wirbel, B. Stech and M. Bauer, Z. Phys. {\bf C29} (1985) 637.

\bibitem{Form2} M. Bauer, B. Stech and M. Wirbel, Z. Phys. {\bf C34} (1987) 103.



\bibitem{PDG} W. M. Yao et al, J. Phys. {\bf G33} (2006) 1.


\bibitem{decayCP1} M. Artuso et al, Phys. Rev. Lett. {\bf 95} (2005)
251801.
\bibitem{decayCP2} G. Bonvicini et al,  Phys. Rev. {\bf D70} (2004)
112004.

\bibitem{decayCPs}T. K. Pedlar  et al, Phys. Rev. {\bf D76} (2007) 072002.

\bibitem{Narison0807} S. Narison, Phys. Lett. {\bf B668} (2008) 308.

\bibitem{HQEFT97} R. Casalbuoni, A. Deandrea, N. Di Bartolomeo, R. Gatto, F. Feruglio and
 G. Nardulli, Phys. Rept. {\bf 281} (1997) 145.

  \bibitem{VMDgV} M. Bando, T. Kugo and K. Yamawaki,  Phys. Rept. {\bf 164} (1988)
217.

\bibitem{WangCCT1}  Z. G. Wang, Nucl. Phys. {\bf A796} (2007) 61.

\bibitem{WangCCT2} Z. G. Wang, Eur. Phys. J. {\bf C52} (2007) 553.


\bibitem{HeavyQuark} M. Neubert, Phys. Rept. {\bf 245} (2004) 259.



\bibitem{3P01} A. Le Yaouanc, L. Oliver, O. Pene and  J. C. Raynal, Phys. Rev. {\bf D8} (1973)
2223.

\bibitem{3P02} A. Le Yaouanc, L. Oliver, O. Pene and J.-C. Raynal, Phys. Rev. {\bf
D9} (1974) 1415.

\end{thebibliography}
\end{document}